\documentclass[aps,prstper,reprint,groupedaddress]{revtex4-2}
\usepackage{graphicx, tabularx, longtable, nicefrac, amsmath,enumerate, supertabular, eurosym}
\usepackage{microtype}
\usepackage{booktabs}
\usepackage{csquotes}
\usepackage[utf8]{inputenc}
\usepackage{booktabs}
\usepackage{hyperref,xcolor}

\begin{document}


\title{Physics task development of prospective  physics teachers using ChatGPT}




\author{Stefan K{\"u}chemann$^1$}
\email{mailto: s.kuechemann@lmu.de}
\author{Steffen Steinert$^1$}
\author{Natalia Revenga$^1$}
\author{Matthias Schweinberger$^1$}
\author{Yavuz Dinc$^1$}
\author{Karina E. Avila$^2$}
\author{Jochen Kuhn$^1$}

\affiliation{$^1$ Chair of Physics Education Research, Faculty of Physics, Ludwig-Maximilians-University Munich, Geschwister-Scholl-Platz 1, 80539 Munich, Germany}
\affiliation{$^2$ Department of Mathematics, RPTU Kaiserslautern-Landau, Paul-Ehrlich-Str. 14, 67663 Kaiserslautern, Germany}


\date{\today}

\begin{abstract}
The recent advancement of large language models presents numerous opportunities for teaching and learning. Despite widespread public debate regarding the use of large language models, empirical research on their opportunities and risks in education remains limited. In this work, we demonstrate the qualities and shortcomings of using ChatGPT 3.5 for physics task development by prospective  teachers. In a randomized controlled trial, 26 prospective  physics teacher students were divided into two groups: the first group used ChatGPT 3.5 to develop text-based physics tasks for four different concepts in the field of kinematics for 10th grade high school students, while the second group used a classical textbook to create tasks for the same concepts and target group. The results indicate no difference in task correctness, but students using the textbook achieved a higher clarity and more frequently embedded their questions in a meaningful context. Both groups adapted the level of task difficulty easily to the target group but struggled strongly with sufficient task specificity, i.e., relevant information to solve the tasks were missing. Students using ChatGPT for problem posing rated high system usability but experienced difficulties with output quality. These results provide insights into the opportunities and pitfalls of using large language models in education.

\end{abstract}


\maketitle

\section{Introduction}

The ability to develop tasks is an essential skill for teachers in schools, particularly for creating formative and summative assessments. High-quality tasks enable teachers to diagnose student conceptual understanding and difficulties, monitor progress, and evaluate the efficacy of pedagogical methods. Utilizing well-designed assessments, teachers can implement timely interventions and support student development, directly impacting student performance. By creating and implementing diverse tasks that accommodate varying learning preferences and prior knowledge levels, teachers can foster inclusive classrooms and account for student diversity, directly impacting educational quality. Consequently, standards for teacher competence in student assessment were established in 1990 \cite{aft9nea}, and the competence was recognized as a relevant component of pedagogical content knowledge \cite{gess2019teacher} and of Classroom Assessment \cite{airasian2001classroom}. Despite its acknowledged relevance, research indicates that teachers in the United States feel inadequately prepared to develop tasks assessing student performance~\cite{mertler1998classroom,stiggins1999you}, despite receiving undergraduate assessment training. This sentiment aligns with findings of limited assessment literacy~\cite{popham2003seeking}. It is therefore essential to identify methods for training prospective and in-service teachers in task development, particularly with digital tools \cite{eyal2012digital}.

In addition to numerous other applications, recent advancements in large language models, such as ChatGPT, present opportunities for the automated generation of assessments utilizing well-crafted prompts \cite{kasneci2023chatgpt}. This raises questions regarding the quality of tasks developed by prospective  teachers using ChatGPT compared to traditional methods, such as textbook assistance. By enabling teachers to specify prompts and task creation criteria, ChatGPT may facilitate the streamlining and enhancement of the task development process.


This study compares the quality of tasks developed by prospective  physics teachers utilizing ChatGPT (intervention group) to those created using a textbook (control group). The research focuses on three primary questions:\\

\begin{itemize}
\item[RQ1:] What is the quality of physics tasks developed by prospective  teachers using ChatGPT in comparison to a textbook? 
\item[RQ2:] What improvements do prospective  physics teachers make to ChatGPT-generated and textbook tasks?
\item[RQ3:] How do physics teachers rate ChatGPT’s usability, perceived usefulness, and output when creating physics tasks? 
\end{itemize}

In the following, we underpin our research questions by summarizing the current state of research on assessments in high school physics, teachers’ ability to create assignments, and the rapidly evolving field of large language models in education. Our analysis will focus on the parameters outlined in Section \ref{subsec:assignments} to determine if using ChatGPT leads to improved quality of physics assessment tasks created by prospective  teachers.


\section{Theoretical background}
\subsection{Assessments in high-school level physics}
\label{subsec:assignments}

Physics education fundamentally  entails the utilization and resolution of various tasks that facilitate the organization of student learning, enable teachers to monitor student progress, and provide a means for performance measurement~\cite{fischer2022physics}. Task resolution is essential for comprehending physics, despite its perception as an onerous prerequisite for written performance measurement~\cite{leisen2001qualitatsteigerung}. Students must engage with a multitude of tasks to successfully understand physics, requiring the differentiation between learning and performance tasks~\cite{fischer2022physics}, as well as an understanding of when errors are permissible~\cite{neumann2013towards, osborne2016development}. Tasks may be presented orally or in written form, with test tasks predominantly written. Text-based assignments necessitate reading proficiency, the application of physical concepts and mathematical procedures, and critical reflection~\cite{villarroel2020using}. 

The process of developing an effective task or adapting it to the varying conditions of a classroom is a creative and iterative endeavor. It involves a cycle of trial and error, followed by revisions to the task.~\cite{fischer2022physics}. Examination tasks should assess individual competencies; multiple-choice tasks are appropriate but may be one-dimensional. Such one-dimensional tasks are characteristic of regular test culture and can be statistically evaluated to provide a clear overview of student competencies~\cite{fischer2022physics}. However, texts utilized in testing must be unambiguous to prevent misinterpretation. Failure in such tasks is typically attributed to text misinterpretation or insufficient knowledge or foundational understanding~\cite{fischer2022physics}. Traditional tasks often neglect student learning processes, impeding the development of resilient concepts~\cite{sinaga2017enhancing}. These tasks primarily assess reading or mathematical competencies rather than physical conceptual thinking. 

The implementation of competency-based learning standards and curricula in countries such as the USA~\cite{calmer2019teaching} and Germany~\cite{ sekretariat2005bildungsstandards, kremer2012assessment} requires further development of the task culture~\cite{leisen2001qualitatsteigerung}. Tasks that focus on student competence and development can make the metastructures of physics content (basic concepts and guiding ideas) and the physics competencies to be learned more transparent by applying them to examples and encouraging reflection~\cite{fischer2022physics}. Competency-based tasks typically feature a strong contextual orientation and are often preceded by text or material containing both relevant and irrelevant information to provide contextual framing. Visual aids such as pictures, graphics, or newspaper clippings may also be used as part of this introduction. Subsequent assignments are given using command verbs~\cite{leisen2006aufgabenkultur} and take into account sub-competencies such as evaluation and communication. These assignments may include means of processing and presentation, aids and hints for processing, information on presentation, and processing and evaluation criteria. The most challenging tasks are open-ended experimental complex problem-solving tasks~\cite{akben2020effects}, which allow for multiple reasonable solutions and may not have a single unambiguous answer. In the case of open-ended tasks, the underlying question must typically be clarified before the problem can be solved.

Tasks are an essential component of physics lessons and associated learning assessments. Despite their routine use, the selection and creation of appropriate tasks should not be underestimated. Tasks must effectively capture intended learning goals, engage a typically diverse learning group, and accommodate individual understanding of physical concepts~\cite{fischer2022physics}.

\subsection{Teachers' ability to create assessments}
\label{sec:teachers_ability}

Extensive research has been conducted on the pedagogical and domain-specific skills required by teachers to effectively facilitate student learning, resulting in the proposal of several models describing fundamental teacher competencies~\cite{Rowland2005,shulman}. Recent studies in mathematics and science indicate that teachers’ pedagogical content knowledge (PCK) significantly impacts student achievement~\cite{GessN}. Shulman emphasizes the importance of PCK, which enables teachers to make subject matter accessible to students by combining subject knowledge and teaching skills~\cite{Kleickmann2012, shulman, shulman87}. Subject knowledge presupposes the possession of problem-solving skills among educators. However, to effectively employ these skills for didactic purposes through selection, adaptation or developmnt, educators must possess a thorough understanding of task didactic analysis~\cite{Edwards2015TheIO}. 


Educators are expected to possess the ability to develop and reformulate problems in order to facilitate meaningful learning environments for students~\cite{Lee2018}. This activity, referred to as problem posing by Silver~\cite{Silver1994}, encompasses both the generation of novel problems and the reformulation of existing tasks. 
Problem posing is crucial for both students and educators, as it fosters creative thinking among students~\cite{Bonotto2013} and provides educators with insight into students’ thought processes and conceptual understanding~\cite{Cai2020}. For instance, educators have successfully identified mathematical misconceptions among students through the use of problem posing tests and student self-posed problems~\cite{Koichu2013}.


The ability of educators to apply and teach problem posing significantly impacts students’ conceptual understanding of problem solving. As such, it is crucial for educators to possess the skills and knowledge necessary to design and reformulate problems in order to facilitate similar learning activities for their students~\cite{Lee2018}. Lowrie (2002) highlights the close correlation between problem posing and problem solving as cognitive processes~\cite{Lowrie2002}. As a result, problem posing constitutes a critical component of educators’ responsibilities, as they are tasked not only with presenting problems to students but also with guiding them towards becoming proficient problem posers in their own right~\cite{Crespo2003}. 


Despite its importance, problem posing presents several challenges. For example, Crespo and Sinclair (2008) found that prospective  teachers experienced difficulty engaging with problem posing due to their lack of familiarity with it in their roles as both students and educators~\cite{Crespo2008}. Additionally, the quality of posed problems is not always high, as demonstrated by Cai and Hwang (2002)~\cite{Cai2002} and Silver and Cai (1996)~\cite{Silver1996cai}. These findings underscore the need for further research into how educators learn to incorporate problem posing into their instruction and how they can improve the quality of their own problem sets in order to effectively integrate them into their teaching~\cite{Cai2020hwang}.


\subsection{Large language models in education}
The ChatGPT system~\cite{openai_chat_gpt_2023} use in this work is based on a Large Language Model (LLM). LLMs are neural networks for natural language processing (NLP) that are trained on extensive text datasets and capable of generating human-like text for a variety of language-related tasks~\cite{kasneci2023chatgpt}. Models such as GPT~\cite{floridi2020gpt}, BERT~\cite{devlin2018bert}, and RoBERTa~\cite{liu2019roberta} have revolutionized the field of NLP and expanded the possibilities for research and applications. In education, these models can be integrated with chatbots to create adaptive and personalized learning experiences for students while supporting educators in their roles~\cite{zhu2020effect, bernius2022machine}. LLMs have been employed to assist students in numerous ways, including content generation~\cite{KASNECI2023102274, raina2022multiple}, improving question-asking skills~\cite{abdelghani2022gpt}, generating code explanations~\cite{sarsa2022automatic}, automating assessments~\cite{jia2021all}, and providing feedback in language learning~\cite{jeon2021chatbot}. Chatbots can serve as conversational partners, supporting students experiencing foreign language anxiety~\cite{bao2019can} or low communication readiness~\cite{tai2020impact}. The integration of LLMs and chatbots can facilitate more engaging learning experiences and aid students in expressing curiosity and comprehending complex concepts~\cite{abdelghani2022gpt}. Research indicates that educators hold positive attitudes towards AI in education~\cite{polak2022teachers}, with factors such as perceived usefulness, ease of use, and trust in AI-based tools influencing their acceptance~\cite{chocarro2023teachers}.

Despite the potential of AI and chatbots in education, several challenges and open questions remain to be addressed. The responsible integration of AI into education will require the collaborative efforts of diverse communities, including educators, researchers, and policymakers~\cite{holmes2020artificial}. 
Further research is required to investigate the effectiveness of LLMs and chatbots in various educational contexts, evaluate their impact on learning outcomes, and examine potential ethical concerns and biases~\cite{kasneci2023chatgpt}. 

The version of ChatGPT evaluated in this study, released on January 30$^{\text{th}}$, demonstrated some limitations. Based on GPT-3.5, this version was unable to process images. Prior to the study, according to its own information, it exhibited several limitations including a lack of common sense knowledge, the potential for biased output, limited conversational context, and difficulty with abstract reasoning and creativity~\cite{chatgpt_2023_limitations_1}.


Upon further inquiry on February 27 regarding the limitations of the ChatGPT version based on GPT-3.5, additional limitations were reported. These included knowledge is limited in time, the potential for ambiguous or unclear responses, the generation of plausible but incorrect or misleading answers, sensitivity to input phrasing, verbosity, the potential for inappropriate content, a lack of common sense, an inability to ask clarifying questions, and limited consideration of longer conversational contexts~\cite{chatgpt_2023_limitations_2}.


After the completion of the study, a version of ChatGPT based on GPT-4 became available for testing~\cite{openai_gpt4_2023, openai2023gpt4}, which provided information about the differences between its limitations and those of the GPT-3.5-based version. The limitations of ChatGPT versions based on GPT-4 and GPT-3.5 are largely similar, as both models share common issues in their design and training. However, GPT-4 can potentially provide more current information than GPT-3.5, provide higher quality responses, handle ambiguous questions more effectively, be better at maintaining context, and have a potentially lower frequency of inappropriate content generated by the model~\cite{chatgpt_2023_limitations_2}. 

It has been announced that GPT-4 will be able to process images~\cite{openai_gpt4_2023}, opening up new possibilities for educational applications.



LLMs and chatbots have the potential to revolutionize education by providing adaptive and personalized learning experiences. These technologies can assist learners in acquiring knowledge and support teachers in their roles, thereby enhancing the engagement and effectiveness of education. However, it is crucial to further investigate their capabilities and address the challenges associated with their integration to ensure their responsible and ethical use in educational settings.


\section{Methods}
\subsection{Participants}
The aim of this study is to examine the potential of chatbots in assisting physics educators in the development of appropriate assessment tasks for high school students. Given that the challenges associated with task creation (see Sec. \ref{sec:teachers_ability}) are often related to experience, the focus of this investigation is on undergraduate prospective  physics teacher students and graduate prospective  physics teachers who do not regularly teach physics.


A total of 26 prospective  physics teachers (13 female and 13 male) from Ludwig-Maximilians-University Munich, with a median age of 23 years, participated in this study. Of these participants, 80\% had one year or less of teaching experience and 95\% of this subgroup (76\% of the total sample) reported having little to no experience in creating assessment tasks. Only 15\% of participants reported having moderate experience in this area, while a mere 7\% claimed to have extensive experience, with teaching experience ranging from 3.5 to 6 years. Participation was voluntary and uncompensated. Prior to the commencement of the study, the local ethics board reviewed and approved to the performance of the study.


\subsection{Design and Materials}
In this study, participants were tasked with developing four conceptual tasks to assess common concepts in Newtonian physics for 10th-grade German high school students (aged 15-16 years). These concepts included the relationship between velocity and acceleration and the first, second, and third laws of motion.
 These concepts were selected due to their fundamental importance in the high school physics curriculum. Half of the participants used ChatGTP 3.5 \cite{openai_chat_gpt_2023} (based on the January 30$^{\text{th}}$ version of ChatGTP 3.5) as a support to create physics tasks (intervention group), while the other half had access to a digital standard high school physics textbook as a support to create the physics tasks (control group)\cite{clickandstudy}. Physics textbooks for high schools often contain tasks that teachers may use in class. However, to prepare an exam, teachers often need to modify textbook tasks to test students’ problem-solving skills and conceptual understanding without relying on memorization of the results. To account for this practice and to allow for a fair comparison between the two groups, participants in the control group were asked to make significant changes to the textbook tasks (to change more than just the given values). Moreover, the participants in the intervention group did not have access to the physics textbook and the control group did not have access to ChatGPT. 

Prior to task creation, intervention group students were asked to enter three given prompts into ChatGPT
(1. “Create a question about the book Goethe's Faust.”, 2. “Create a question on Hermann Hesse's book Demian on the role of Demian for 11th grade students in a high school.", 3. "Create a multiple-choice question on the importance of Quidditch to Harry Potter in the book Harry Potter and the Sorcerer's Stone.") 
to familiarize themselves with ChatGPT’s sensitivity and flexibility in responding. After completing these prompts, students were given the opportunity to experiment with their own inputs before proceeding to the conceptual task creation phase. 


In each group, participants were instructed to provide both their own created or adapted questions and the original questions from either the textbook or ChatGPT that inspired their task development or adaptation, if applicable.
This serves two purposes: first, to ensure that participants in the control group do not simply copy questions directly from the textbook without making modifications. Such a result would not be valid for our study, as our aim is to evaluate participants' ability to create original assessment tasks. Second, we aim to observe to what extent participants modify the tasks suggested by ChatGTP. This enables us to evaluate both ChatGPT's ability to generate high-quality assessment tasks, and the total workload required for the educator when using the tool. The chat-history has been saved to verify students' inputs.

As an educator’s ability to create suitable assessments is influenced by their own conceptual understanding of physics, participants’ prior knowledge was evaluated using the half-length Force Concept Inventory (FCI) version 2~\cite{han2015dividing}.
The test was selected because it covers the aforementioned concepts for which students developed the tasks. A good understanding of the subject matter might be crucial to compensate for potential limitations of ChatGTP in generating consistent tasks~\cite{chatgpt_2023_limitations_1, chatgpt_2023_limitations_2}. It should be noted that standard textbooks typically do not suffer from this limitation, as they undergo regular quality checks and revisions. However, even in this case, it is also essential for educators to possess a strong grasp of the assessed concepts and task quality when it comes to modifying textbook tasks for final exams.

Finally, given that large language models are a relatively new tool and may be associated with user-related challenges, a usability survey was conducted with intervention group participants. This survey included ten questions from the System Usability Scale (SUS)~\cite{schaefer2009effects} and six questions regarding perceived usefulness and output quality from the Technology Acceptance Model 2 (TAM2)~\cite{tam2}. Additionally, an extra question was included to assess whether unexpected slowdowns in ChatGPT’s average response time affected the user experience (see Supplementary Material). Each question was answered on a 5-point Likert scale: Do not agree at all, do not agree, neutral, agree, fully agree. At the end of the study, participants were asked to provide their demographic data.


\subsection{Qualitative data analysis of developed tasks}
In total, participants developed $N=$103 assessment tasks ($N=$51 in the intervention group and $N=$52 in the control group). Based on previous literature, eight categories were identified for evaluating the developed assessment tasks: 
specificity (scale: 0$=$does not apply; 1$=$applies), clarity (scale 0/1), correctness (scale 0/1), not misleading (scale 0/1), adequate difficulty (scale 0/1), context (scale 0/1), relevance to map the target concept (5-point-Likert scale: completely irrelevant, rather irrelevant, about the same number of irrelevant as relevant aspects, rather relevant, completely relevant), and overall quality (5-point-Likert scale: very low, rather low, medium, rather high, very high) (see supplementary material for an explanation of these categories).

Two independent raters with over six years of experience in teaching prospective  physics teachers, conducting physics education research, and developing conceptual questions for high school and university-level physics rated each task according to these eight categories. The interrater reliability Cohen’s $\kappa$ between the two raters was determined for the first six categories (see Tab. \ref{tab:Interrater}), while the other two categories were averaged (see Results).

\begin{table}[]
    \centering
    \begin{tabular}{l c}
    \hline \hline
 Category	& Cohen's Kappa \\ \hline 
Specificity	~~~~~~~& 0.43 \\
Clarity	~~~~~~~& 0.52 \\
Correctness & 0.56 \\
Not Misleading	& 0.35 \\
Adequate Difficulty	~~~~~~~& 0.76 \\
Context	& 0.66 \\ \hline
Average & 0.55\\ \hline \hline
    \end{tabular}
    \caption{Interrater reliability of the rating of developed tasks.}
    \label{tab:Interrater}
\end{table}
According to Landis and Koch, these values of Cohen’s $\kappa$ can be interpreted as indicating fair agreement ($\kappa=0.35$) in the lowest case of whether a task’s phrasing is misleading, up to substantial agreement ($\kappa=0.76$) in the category of whether a task has adequate difficulty for a 10$^\mathrm{th}$ grade high school physics class \cite{landis1977measurement}. On average, we find moderate agreement between the two raters ($\kappa=0.55$). Disagreements in ratings were resolved through discussion

\section{Results}
\subsection{Quality of generated tasks}
\begin{figure}
\centering
\includegraphics[width=\linewidth]{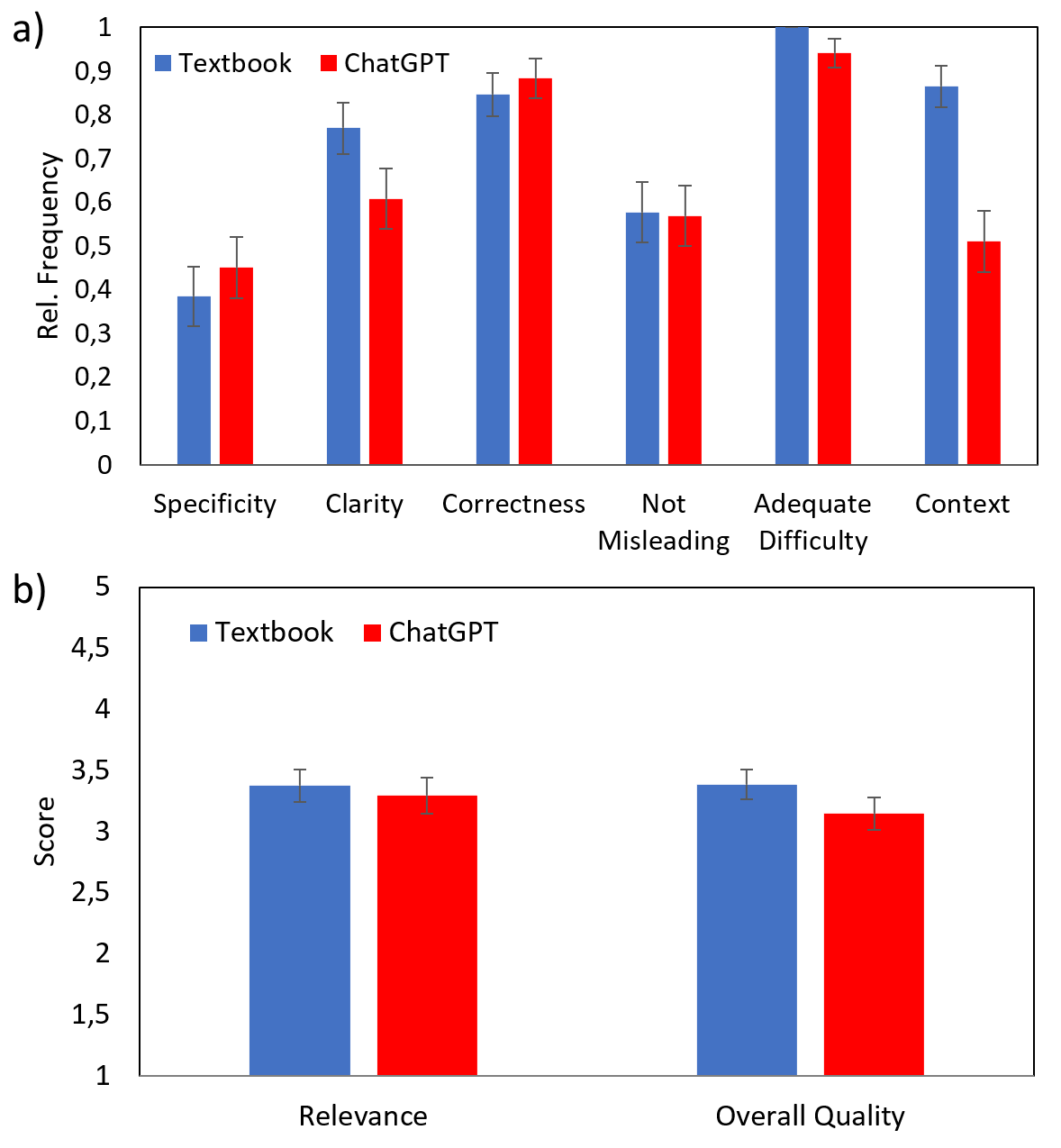}
\caption{Differences between the control group (textbook) and intervention group (ChatGPT) in (a) six categories rated on a scale from 0 (does not apply) to 1 (applies), and (b) categories rated on a Likert scale from 1 (very low) to 5 (very high). 
}
\label{Fig1}
\end{figure}

After resolving disagreements, we compared the differences between the two groups across the eight categories (see Fig. \ref{Fig1}). A $t$-test revealed no significant difference in task specificity, correctness, whether phrasing was misleading, or whether tasks had adequate difficulty (each~$p>0.05$). It is interesting to note that both groups achieved (nearly) perfect scores in task adequacy for the target group and very high levels of task correctness. In contrast, both groups achieved only low values in task specificity of around 0.4, indicating that students in both groups had difficulty providing sufficient information for tasks to be solvable. Apart from that, we found significant differences in task clarity ($p$=0.04) with a small effect size of Cohen’s $d=0.38$, and task context ($p=6\cdot 10^{-5}$) with a large effect size of $d=1.04$ between the two groups. In both cases, students who worked with the textbook achieved higher scores.

\begin{table*}[ht!]
    \centering
    \begin{tabular}{p{3cm} p{7cm} @{\hspace{.5cm}} p{7cm} }
     \hline \hline \\[-0.1cm]
    Aspect & ChatGPT 3.5 & Textbook \vspace{1mm}\\ \hline \\[-0.3cm]
     Overall high quality & A car accelerates from an initial speed of 20~m/s to a final speed of 60 m/s in a period of 10 seconds. Calculate the acceleration of the car within this time period assuming that the acceleration is constant. & Justify why space debris in Earth orbit can be very dangerous for a space station like the ISS. Argue with Newton's first law.\\\\
    Lack of Specicifity & A car accelerates from a standstill to a speed of 100~km/h.
     \begin{enumerate}
         \item Explain the difference between the physical quantities acceleration and velocity and explain the units. 
         \item Calculate the acceleration of the car in m/{s}$^2$.
     \end{enumerate} & A body moves frictionless with the velocity v$_0$. What is the velocity of the body after 10~s? Justify your decision using Newton's laws.  \\\\
    Missing context & How does Newton's 1st law describe the behavior of a body in terms of its motion or rest? & How does the average velocity of a constant positive acceleration process from a standstill relate to the terminal velocity? What tools can you use to do this? \\ \hline \hline
    \end{tabular}
    \caption{Specific examples of tasks generated using ChatGPT 3.5 or a textbook that demonstrate high quality, lack of specificity, or missing context.}
    \label{tab:2a}
\end{table*}
Table~\ref{tab:2a} shows some examples of tasks developed by students in the ChatGPT and in the textbook group. In the first example, the tasks from both groups reached an overall high quality in the ratings in the eight categories, e.g., they are correct, clear, concise, embedded in a meaningful context, and map the target concept. In the second example, the task in the ChatGPT group misses the time required for a car to accelerate from a standstill to a speed of 100 km/h, thus it is not sufficiently specified as the acceleration of the car can be determined. Similarly, in the second task in the textbook group, the participant mentioned that there is no friction occurring but fails to specify if another force is acting on the body. In the third example, both groups’ tasks lack context.

Furthermore, we investigated whether the scores achieved by students in each group were related to their prior knowledge. The intervention group participants achieved an average FCI score of $0.62\pm 0.05$, while control group participants achieved an average FCI score of $0.69\pm 0.05$ (overall average: $0.65\pm 0.05$), but the difference between the two groups was not significant ($p>0.05$). To this end, we performed a logistic regression between students’ FCI scores and the ratings of the tasks they created in the first six categories, which were rated on a scale of 0 and 1. We found no significant relationship between FCI scores and ratings in the first six categories. Additionally, we performed a linear regression between students’ FCI scores and the ratings of the last two categories, which were rated on a scale from 1 to 5. We also found no linear relationship between these two quantities. Consequently, in this study, prior knowledge had no significant influence on the quality of the tasks created.


\subsection{Adaptations to textbook and ChatGPT-generated tasks}

In addition to evaluating the quality of tasks in the eight categories, we also assessed the changes students made to textbook tasks compared to ChatGPT-generated tasks. We evaluated whether these changes led to an improvement (+1) or a decrease (-1) in task quality. Overall, one participant in the ChatGPT group did not use ChatGPT as a support tool for one task (2\% of all tasks in this group) and instead created the task without any support. In contrast, 42 tasks (81\% of all tasks in this group) were developed by participants in the textbook group without using the textbook as a resource. Consequently, participants in the textbook group used the textbook for ten tasks (19\% of all tasks in this group) and adapted the given questions. In comparison, participants in the ChatGPT group used ChatGPT for 50 tasks (98\% of all tasks in this group) and adapted ChatGPT tasks in 12 cases (24\% of all tasks in this group). This means that students in the ChatGPT group used 38 tasks (75\% of all tasks in this group) as provided by ChatGPT.

\begin{figure}[ht!]
\centering
\includegraphics[width=\linewidth]{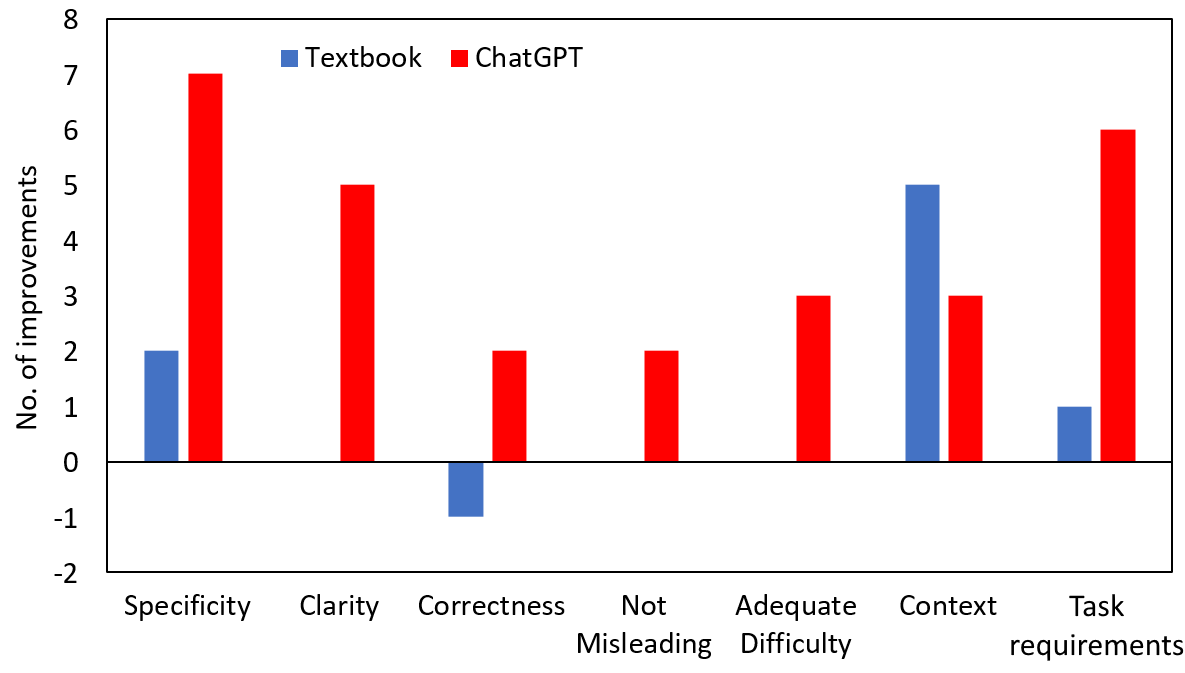}
\caption{Number of improvements made by students to textbook or ChatGPT-generated tasks within the seven categories. Negative values indicate that a participant's adaptation decreased the task quality. 
}
\label{Fig2}
\end{figure}

Fig.~\ref{Fig2} shows the number of improvements students made in seven categories. A positive value indicates that more changes led to task improvement, while a negative value indicates that more changes led to a decrease in the task quality. The first six categories were identical to those used in the quality rating, with the exception of the category regarding relevance to map the concept, which was omitted. The overall quality was also omitted and a new category reflecting changes in task requirements was added. In this new category, we evaluated whether prospective  physics teachers adapted the type of the task or what the task required the students to do. Table~\ref{tab:2} presents four instances of such task modifications made by participants using either ChatGPT-generated tasks or tasks extracted from the textbook.
\begin{table*}[ht!]
    \centering
    \begin{tabular}{c p{2cm} p{6.5cm} @{\hspace{.5cm}} p{6.5cm} }
     \hline \hline \\[-0.1cm]
    Example No. & Group & ChatGPT-generated or textbook task & Adapted task by the participant \vspace{1mm}\\ \hline \\[-0.3cm]
     1 & ChatGPT  & A spaceship has a mass of 2000~kg and is in orbit around the planet X. The gravitational force of planet X on the spacecraft is 20000~N. 
     \begin{enumerate}
         \item Calculate the acceleration of the spacecraft with respect to planet X. 
         \item What is the velocity of the spacecraft in orbit?
         \item What is the size of the trajectory of the spacecraft when the speed of the spacecraft is changed?
     \end{enumerate} & A spaceship has a mass of 2000~kg and is in orbit around the planet X. The gravitational force of planet X on the spacecraft is 20000~N. 
     \begin{enumerate}
         \item Calculate the acceleration of the spacecraft with respect to planet X. 
         \item What is the velocity of the spacecraft in orbit?
         \item Describe the trajectory of the spacecraft when the speed of the spacecraft is changed.
     \end{enumerate}\\ 
    2 & ChatGPT & A car accelerates from an initial speed of 20~m/s to a final speed of 60~m/s in 10 seconds. Calculate the acceleration of the car. & A car accelerates from an initial speed of 20~m/s to a final speed~of 60 m/s in 10 seconds. Calculate the acceleration of the car within this time assuming that the acceleration is constant. \\\\
    3 & Textbook & The reason for the propulsion of a rocket, as for an inflated balloon that you release, is the interaction principle. Explain the statement. & In order for a Boing 747 to get enough thrust to fly, Newton's Law of Interaction works much like letting go of an inflated ball of air. Explain this principle in the context of the Boeing 747 flying (or taking off). \\\\
    4 & Textbook & Selma (m1=60~kg) takes a running jump onto a stationary sled (m2=12~kg). Both then continue to travel at u=2.5~m/s.
    \begin{enumerate}
        \item Calculate the velocity of Selma, with which she landed on the sled.
        \item Prove that the kinetic energy is not conserved during the process.
    \end{enumerate}
    & Selma (m1=60~kg) takes a running jump onto a moving sled (m2=12~kg, v2=2~m/s). Both continue to travel together.
    \begin{enumerate}
    \item Calculate the joint final velocity, assuming no friction and no gradient.
    \item Calculate the joint momentum at impact.
    \end{enumerate}\\ \hline \hline
    \end{tabular}
    \caption{Examples of participants' modifications of the tasks that were either developed by ChatGPT or extracted from the textbook.}
    \label{tab:2}
\end{table*}



In Table~\ref{tab:2}, example 1 illustrates an adaptation of the task type by a participant in the ChatGPT group. In this instance, the prospective  physics teacher failed to recognize that the velocity of the spacecraft in part 2 of the task was indeterminable given the available information. Nevertheless, the participant shifted their focus to modifying the task requirements in part 3 from a quantitative to a descriptive task. Fig.~\ref{Fig2} shows that participants working with ChatGPT made a greater number of improvements compared to those in the textbook group. This is expected as participants in the textbook group primarily developed tasks independently of the textbook, resulting in no recorded changes to textbook tasks. In contrast, students in the ChatGPT group primarily altered task specificity and requirements. For instance, in Table~\ref{tab:2}, example 2 demonstrates a modification of task specificity by a participant in the ChatGPT group after ChatGPT generated the task. 
In this instance, it is impossible to determine the acceleration of the car due to the ambiguity surrounding whether its acceleration is constant. The participant recognized this limitation and subsequently modified the task accordingly, resulting in a well-defined and specific final task.








The textbook group primarily focused on modifying the context of the tasks. In Table~\ref{tab:2}, Example 3 illustrates that both tasks are correct with only a minor alteration in context. However, there were instances where participants made incorrect changes to textbook tasks. For instance, in Table~\ref{tab:2}, Example 4 demonstrates a failure to address the fact that the task does not focus on the relationship between velocity and acceleration, but rather on the conservation of momentum during an inelastic collision. Additionally, the participant neglected to include Selma’s initial velocity. As a result, the final velocity in part 1) and the joint momentum in part 2) of the task can no longer be determined.





\subsection{Usability of ChatGPT 3.5 to create physics tasks}
\begin{figure}
\centering
\includegraphics[width=\linewidth]{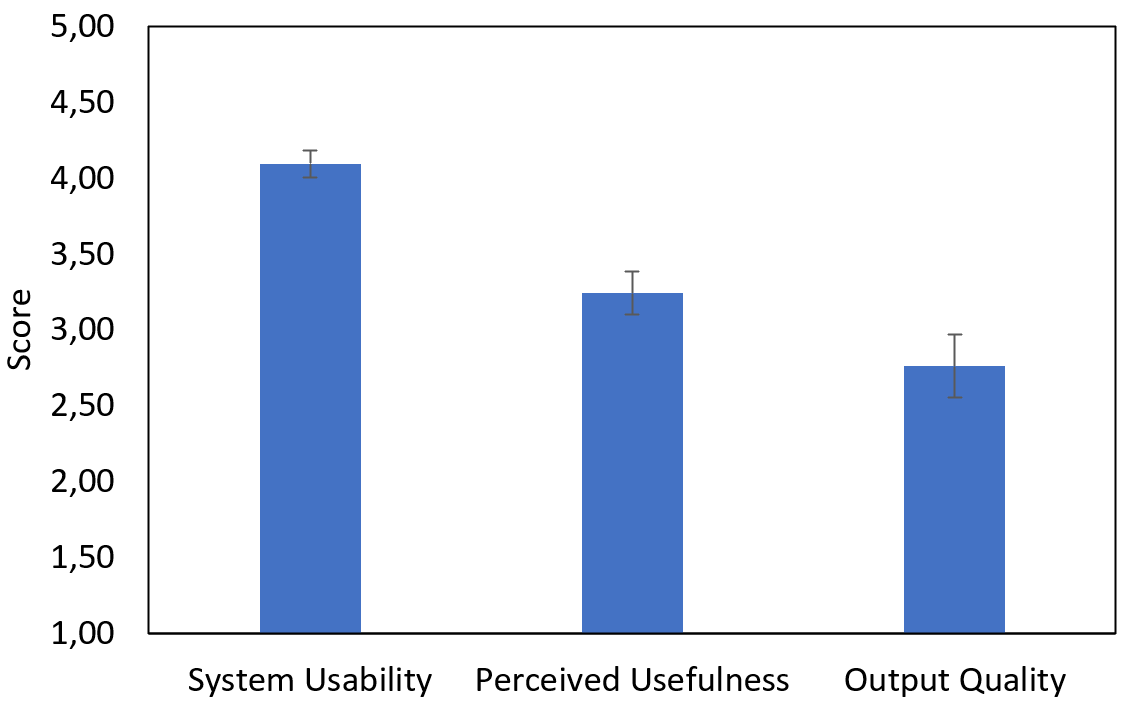}
\label{Fig3}
\caption{Rating of participants regarding the ChatGPT's usability, its perceived usefulness, and its output quality on a 5-point Likert scale: 1=strongly disagree, 2= disagree, 3=neutral, 4=agree, 5=strongly agree. For the analysis, the items with negative phrasings have been switched, so that high values reflect a positive statement of the students.}
\end{figure}

To evaluate how participants perceived the output they received from ChatGPT and the ease or difficulty of receiving a response from the system of sufficient quality, we asked participants to rate the usability, perceived usefulness, and output (Fig.~\ref{Fig3}). On average, participants rated system usability at 4.1, indicating agreement that the system is easy to use, functions are well-integrated, and they were confident in using it. Regarding perceived usefulness in developing physics tasks, participants were neutral (average value 3.24). This implies that they did not feel more productive or effective but were also not slowed down in their efforts to develop physics tasks. This scale also included a question about whether the time until receiving a response from ChatGPT complicated system use. On average, students did not agree (average value 2.09) that the response time complicated system use. Additionally, participants rated ChatGPT’s output quality as neutral but with the lowest average value of 2.76, indicating that participants tended to have difficulties with the system’s output.

\section{Discussion}
The aim of the manuscript was the study of three research questions. The first research question targeted the quality of tasks generated by participants using ChatGPT compared to tasks created by students using a textbook. Both groups achieved very high task correctness, exceeding the average score in the concept test. Assuming that the difficulty of generating a correct assessment task is comparable to the average difficulty in the half-length force concept inventory, this finding suggests that ChatGPT is able to compensate for conceptual difficulties that prospective physics teachers may have, reducing their likelihood of translating to assessment tasks.


Apart from that, it was noticeable that both groups achieved low task specificity, indicating that information was missing to complete the task. This difficulty among prospective physics teachers could not be compensated by either ChatGPT 3.5 or the textbook. It would be interesting to see whether future versions of large language models can overcome this shortcoming or how it can be addressed through learning interventions for prospective physics teachers or other support methods. Additionally, tasks created by participants using ChatGPT were significantly less frequently embedded in the context. However, some tasks generated by ChatGPT had context if participants explicitly prompted it. Therefore, the low score in the ChatGPT group was not caused by the system but rather by students’ insensitivity to the value of the context in a physics assessment task for 10th grade high school students. Students in the textbook group achieved a high score in this category regardless of whether they modified a given textbook task or developed a physics task without inspiration from a textbook task.

The second research question focused on the number and type of changes students made to ChatGPT-generated and textbook tasks. It is relevant to note that 81\% of tasks in the textbook group were not developed based on a given textbook task but were instead created by students without using this resource. This observation suggests that it was less effort for students in this group to include tasks from memory or their own efforts rather than going through the tasks in the textbook. Such behavior likely depends on students’ familiarity with the textbook and how many assessment tasks in the same context they had already created in which they implemented their ideas. The results show that some participants were aware of low specificity in tasks created by ChatGPT and improved it, but most participants did not have this awareness.

Furthermore, the results demonstrate that some students modified pre-existing textbook tasks that resulted in incorrect tasks. Although this observation occurred only in three tasks (6\% of all tasks in this group), and the correctness of all tasks in this group was comparable to the ChatGPT group, such a shortcoming may potentially be compensated by additional use of large language models to solve a given task before giving it to students.

The third research question addressed the usability, perceived usefulness, and quality of ChatGPT’s output. We found that participants rated the usability of ChatGPT 3.5 high, but judged the quality of physics tasks generated by ChatGPT slightly below a neutral level. It would be interesting to see if the judgment of output quality improves with future versions of ChatGPT and whether this judgment would change if students received training to create prompts for ChatGPT to generate physics tasks.

\section{Conclusion}
In this study, we evaluated the quality of tasks developed by prospective physics teachers using ChatGPT 3.5 compared to using a textbook. The correctness, the frequency of tasks with an adequate difficulty, and the overall quality of the tasks created by ChatGPT were comparable to those created by participants who could use a textbook. This is a remarkable achievement for a large language model, as developing physics assessment tasks using textbooks is the current state-of-the-art and requires teachers to have assessment knowledge, conceptual knowledge, and problem-solving skills \cite{brookhart2011educational}. At the same time, the field of large language models are currently quickly evolving and they are likely to become more proficient in such discipline-specific exercises as task development soon.

We found that some shortcomings in tasks, such as specificity, could not be compensated by either the textbook or ChatGPT 3.5, and other aspects needed to be explicitly prompted by the user of ChatGPT, such as the context of the task. 

In summary, we demonstrated that advanced large language models such as ChatGPT 3.5 can effectively support physics teachers in their common practice of developing assessment tasks. It would be beneficial to include specific training for physics teachers in education programs to design prompts for effective use of large language models.

\bibliography{bibliography}

\end{document}